%% file: paper_clean.tex
\documentclass[twocolumn,10pt,amsmath,amssymb,aps,prl]{revtex4-2}
\usepackage{graphicx}
\usepackage{dcolumn}
\usepackage{bm}
\usepackage{siunitx}
\usepackage{fancyhdr}
\usepackage{subfig}
\usepackage{amsmath}
\usepackage{amssymb}
\usepackage{amsthm}
\usepackage{algorithmic, algorithm}
\usepackage{url}
\usepackage{caption}
\usepackage{xcolor}
\usepackage{hyperref}
\usepackage{ulem}

\begin{document}
\title{Exchange-Symmetrized Qudit Bell Bases and Bell-State Distinguishability}

\author{Oscar Scholin}
\affiliation{Cavendish Laboratory, University of Cambridge, JJ Thomson Ave, Cambridge CB3 0HE, U.K.}
\affiliation{Department of Physics, Pomona College, 610 N College Ave, Claremont, CA 91711 U.S.A.}
\email{orsa2020@mymail.pomona.edu}

\author{Theresa W. Lynn}
\affiliation{Department of Physics, Harvey Mudd College, 301 Platt Blvd., Claremont, CA 91711, U.S.A.}
\email{lynn@g.hmc.edu}

\date{\today}

\begin{abstract}

Entanglement of qudit pairs, with single particle Hilbert space dimension $d$, has important potential for quantum information processing, with applications in cryptography, algorithms, and error correction. For a pair of qudits of arbitrary even dimension $d$, we introduce a generalized Bell basis with definite symmetry under exchange of internal states between the two particles. We show that no complete exchange-symmetrized basis can exist for odd $d$. This framework extends prior work on exchange-symmetrized hyperentangled qubit bases, where $d$ is a power of two. For our exchange-symmetrized basis we show that measurement devices restricted to linear evolution and local measurement (LELM) can unambiguously distinguish $2d-1$ qudit Bell states for any even $d$. This achieves the upper bound in general for reliable Bell-state distinguishability via LELM and augments previously known limits for $d = 2^n$ and $d=3$.  This result is relevant to near-term realizations of quantum communication protocols.
\end{abstract}

\maketitle

\noindent  \textit{Introduction---}Entanglement is a key resource in many quantum algorithms and communication protocols, from  use  in prime factorization and machine learning \cite{shor_polynomial-time_1999, kwiat_grovers_2000, harrow_quantum_2009, lloyd_universal_1996, cai_entanglement-based_2015} to quantum secret sharing \cite{scarani_security_2009, senthoor_communication_2019, kogias_unconditional_2017}. In many of these examples, entangled qubits are used, but entangled particles with $d>2$ mutually orthogonal states each, or qudits, are the basis for some computational and cryptographic models \cite{luo_universal_2014, ranade_symmetric_2009}. Although qudits tend to exhibit larger error channels than qubits \cite{otten_impacts_2021}, qudits offer several advantages over qubits including increased density of information (scaling as $d^n$ rather than $2^n$ for $n$ particles), lower decoherence in the underlying physical systems \cite{jankovic_noisy_2024, wang_qudits_2020}, and the ability to encode redundancy in additional dimensions \cite{jankovic_noisy_2024, chiesa_embedded_2021, petiziol_counteracting_2021}. 

Experimental demonstrations of qudit-based algorithms and communication schemes include a variety of quantum communication applications \cite{wang_proof--principle_2018, han_proof--principle_2021, liu_experimental_2012, gaertner_experimental_2007}, as well as quantum algorithms \cite{lim_experimental_2024, raissi_deterministic_2024, meng_experimental_2024,liu_performing_2023}. Central to many of these protocols is measurement in an entangled basis, which necessarily involves unentangling logic gates for complete, deterministic results. These devices are challenging to implement experimentally, especially for photonic states, so many realizations instead use only linear evolution and local measurement (LELM), which can perform only a non-complete Bell state measurement \cite{lutkenhaus_bell_1999, vaidman_methods_1999}. In an LELM device, each particle's evolution through the apparatus is independent of the other particle's presence or state and local measurements are made. LELM devices have been used experimentally to implement quantum teleportation and dense coding protocols in qubits \cite{pan_experimental_1998, mattle_dense_1996, barreiro_beating_2008, bouwmeester_experimental_1997}, teleportation in qutrits \cite{jo_enhanced_2019}, and
dense coding in ququarts \cite{hu_beating_2018}. Given both the relevance of qudit platforms and the use of LELM to enable useful quantum communication protocols experimentally, it is worthwhile to explore the structure of general Bell bases that have properties that can be leveraged for Bell state measurement. 

In this Letter, we present a general complete basis of fully-entangled bipartite qudit states, i.e. a qudit Bell basis, with definite symmetry under particle internal state exchange for any even single particle dimension $d$. We further show that no complete symmetrized basis can exist for odd $d$. Previously a complete symmetrized basis has been presented only for $d = 2^n$ \cite{pisenti_distinguishability_2011, sych_complete_2009}. Our basis allows us to determine the maximum number of general Bell states that can be unambiguously distinguished by an LELM device when $d$ is any even number: $2d-1$. This result joins previously known upper limits for $d=2^n$ \cite{pisenti_distinguishability_2011} and $d=3$ \cite{leslie_maximal_2019}. 

We note that the particle exchange we consider here is not the fundamental exchange symmetry of indistinguishable particles, but rather a symmetry with respect to exchange of internal states between the two entangled particles or subsystems; these two entities are already distinct from each other in an external, typically spatial, degree of freedom. Thus the qudit Bell basis we present here is composed of eigenstates of the SWAP operator with eigenvalues $\pm1$.

 Beyond its applications to qudit Bell state measurement without conditional logic, our basis may also have relevance to extending other quantum information protocols whose qubit versions rely on exchange symmetry of the qubit Bell states. For example, such exchange symmetry plays an important role in numerous quantum communication protocols \cite{eifler_non-local_2020, rodrigues_nonlocal_2020, grosshans_multicopy_2024} and quantum algorithms \cite{jordan_permutational_2010, havlicek_quantum_2018, zheng_speeding_2023, zheng_super-exponential_2022, anderson_symmetric_2017, jacobs_space_2024, divincenzo_universal_2000, freedman_symmetry_2021, rudolph_two-qubit_2023}.


\noindent  \textit{Background---}Consider an arbitrary pure state of a qudit pair, with the general form \cite{sych_complete_2009}:
\begin{equation}
 |\Psi\rangle = \sum_{i, j=0}^{d-1} \chi_{i j} |i\rangle_L|j\rangle_R, 
 \label{eqn:general_pbis}
\end{equation}
where the $L$ (left) and $R$ (right) denote the individual particles distinguished by a spatial degree of freedom: that is, by being stored in distinct memory locations or transmitted along two different communication channels.

Such a pure state is fully entangled if its reduced density matrix (with respect to either subspace) is proportional to the identity \cite{nielsen_quantum_2000}. A complete basis composed of fully entangled states is known as a generalized Bell basis, after the $d=2$ basis of four fully entangled, mutually orthogonal qubit states. The most commonly used qudit Bell basis \cite{bennett_teleporting_1993} defines a Bell state in terms of two parameters, $n,m$:
\begin{equation}
 |\Psi_{nm}\rangle = \frac{1}{\sqrt{d}} \sum_{k=0}^{d-1} e^{2\pi i n k /d } |k\rangle_L | (k + m) \operatorname{mod} d\rangle_R.
 \label{eqn:generalbell}
\end{equation}
Almost all references dealing with generalized Bell state analysis have used this definition, up to a difference in using $(k - m) \operatorname{mod} d$ for the right ket state value or swapping the internal states of the left and right particles \cite{wang_quantum_2008, klimov_mutually_2009, leslie_maximal_2019, hashimoto_simple_2021, yang_local_2017, wu_local_2018, wang_one-way_2019, li_local_2022, karimipour_quantum_2002, tian_classification_2016}. The only exception \cite{horoshko_quantum_2019} introduced a slight variation by mixing the two parameters in the phases: 
\begin{equation}
 |\Psi_{nm}\rangle = \frac{1}{\sqrt{d}} \sum_{k=0}^{d-1} e^{-2\pi i (m - k)n / d} |k\rangle_L | ( m - k ) \operatorname{mod} d \rangle_R.
\end{equation}

For $d=2$, Equation \ref{eqn:generalbell} yields the canonical Bell states which have the special property of being explicitly symmetric or antisymmetric (hence referred to as symmetrized) under the exchange of the left and right particle states. However, for general $d$ the resultant states are not eigenstates of exchange \cite{sych_complete_2009}. Previous work \cite{sych_complete_2009, pisenti_distinguishability_2011} has employed, for $d = 2^n$, a basis of fully entangled mutually orthogonal states by taking explicit tensor products of the $d=2$ Bell states: that is, by constructing the higher-dimensonal basis as hyperentanglement in $n$ qubit degrees of freedom. We extend this work by demonstrating a symmetrized Bell basis for any even $d$, independent of a hyperentangled framework. 

\textit{Symmetric Basis for Even Dimension---}To aid in the construction of this basis, we establish some notation that builds naturally on the canonical $d=2$ Bell states:
\begin{align}
 |\Phi^\pm\rangle &= \frac{1}{\sqrt{2}}(|0\rangle_L |0\rangle_R \pm |1\rangle_L |1\rangle_R), \nonumber\\
 |\Psi^\pm\rangle &= \frac{1}{\sqrt{2}}(|0\rangle_L |1\rangle_R \pm |1\rangle_L |0\rangle_R).
\label{eqn:qubitbell}
\end{align}
Notice that the use of $|\Phi\rangle$ versus $|\Psi\rangle$ in labeling the states takes on the role of $m$ in Equation \ref{eqn:generalbell}; we refer to these as specifying the \textit{correlation class} of a Bell state, since they identify a particular set of pairings between single-particle basis states for the left and right particles. The $\pm$ in Equation \ref{eqn:qubitbell} takes on the role of $n$ in Equation \ref{eqn:generalbell}; we refer to these as specifying the \textit{phase class} of a Bell state since they determine the relative phase between the terms. Based on this correspondence we rename $n, m$ respectively in Equation \ref{eqn:generalbell} as $p, c$ to designate sets of Bell states that share a set of correlations $c$ and phases $p$ \cite{leslie_maximal_2019}. 

By analogy with Equation \ref{eqn:generalbell}, we consider a basis composed of Bell states of the form
\begin{equation}
\label{eqn:general_bell}
 |\Psi_c^p\rangle = \frac{1}{\sqrt{d}} \sum_{k=0}^{d-1} e^{i\phi_{k,p}} |k\rangle_L | \pi_{c}(k) \rangle_R,
\end{equation}
for $p, c \in [0, d-1]$, where $\pi_{c}$ is a permutation of the single-particle basis states. Individual states of this form are inherently fully entangled, so permutations $\pi_{c}$ and phase shifts $\phi_{k,p}$, as specified in Equation \ref{eqn:general_bell}, must be selected to construct $d^2$ states that are mutually orthogonal and symmetrized with respect to exchange of states between the left and right particles.

\noindent  \textit{Correlation classes for exchange symmetry---}To start the construction of the exchange-symmetrized Bell basis, we must show we can choose a set of permutations $\{\pi_c\}$ in Equation \ref{eqn:general_bell} in a way that allows each Bell state to be symmetric or anti-symmetric. Each permutation $\pi_c$ leads to a set of two-particle basis states $\{|k \rangle_L |\pi_c(k)\rangle_R\}$ which appear in $|\Psi_c^p\rangle$. We require these sets to satisfy the following conditions: 

\begin{enumerate}
 \item If $|s_{jc} \rangle_L | t_{jc} \rangle_R$ is included in a set, so is $|t_{jc}\rangle_L | s_{jc} \rangle_R$, to allow for symmetry with respect to particle exchange. 
 \item No $|s_{jc} \rangle_L | t_{jc} \rangle_R$ may appear in multiple sets, so the Bell states in different correlation classes contain non-overlapping two-particle basis states and are thus orthogonal to each other.
\end{enumerate}

The correlation class arising from the trivial permutation, $\pi_0(k)=k$, certainly satisfies the first condition since it involves the set of two-particle basis states $\{|0\rangle_L |0\rangle_R,\dots,|d-1\rangle_L|d-1\rangle_R\}$. 

We can always find a set of $d-1$ additional correlation classes since the problem is equivalent to setting up a round-robin tournament of an even number of players. In particular, consider a group of $d$ players representing the single particle states from $0 \text{ to } d-1$. Each ``round" in the tournament represents a distinct set of $d/2$ pairs of players. It is well known that a $(d-1)$-round complete and non-repeating set of pairings can be achieved for any even $d$ \cite{lucas_recreations_1883}. Beginning from any construction of a $d$-player round-robin tournament, if $(s_{jc},t_{jc})$ play each other in round $c$, the two-particle basis states $|s_{jc}\rangle |t_{jc} \rangle$ and $|t_{jc}\rangle |s_{jc} \rangle$ are included in the Bell states of correlation class $c$. This prescription explicitly enforces our first condition above. Since each pairing can only occur once in the tournament, our second condition is automatically satisfied as well. 

This construction identifies one trivial correlation class ($c=0$) and a further $d-1$ non-trivial correlation classes for use in Equation \ref{eqn:general_bell}.

\noindent  \textit{Phase classes for exchange symmetry---}Our $c\neq0$ candidate Bell states now take the form
\begin{equation}
 |\Psi_{c\neq0}^p\rangle = \frac{1}{\sqrt{d}}\sum_{j=0}^{d/2-1} (e^{i\phi_{j,p}}|s_{jc}\rangle_L | t_{jc}\rangle_R+e^{i\eta_{j,p}}|t_{jc}\rangle_L | s_{jc}\rangle_R).
 \label{eqn:cnotzerorootform}
\end{equation}
To enforce exchange symmetry, we require $e^{i\phi_{j,p}}=\pm e^{i\eta_{j,p}}$, reducing Equation \ref{eqn:cnotzerorootform} to
\begin{equation}
 |\Psi_{c\neq0}^{p}\rangle = \frac{1}{\sqrt{d}}\sum_{j=0}^{d/2-1} e^{i\phi_{j,p}}(|s_{jc}\rangle_L |t_{jc}\rangle_R+(-1)^p|t_{jc}\rangle_L | s_{jc}\rangle_R).
 \label{eqn:cnotzerosymm}
\end{equation}
This formulation also enforces mutual orthogonality of each Bell state pair $|\Psi_{c\neq0}^{p=2\ell}\rangle$ and $|\Psi_{c\neq0}^{p=2\ell+1}\rangle$. To obtain mutual orthogonality of all Bell states within a correlation class, the coefficients $e^{i\phi_{j,p}}$ can now be chosen from the $\frac{d}{2}$th roots of unity as they appear in a discrete Fourier transform (DFT), or: 
\begin{equation}
 \phi_{j,p} = \left \lfloor \frac{p}{2}\right \rfloor \frac{4\pi j}{d}.
 \label{eqn:phijp}
\end{equation}

To obtain a mutually orthogonal set of $c=0$ Bell states there is more freedom in setting the coefficients of each $|k\rangle_L|k\rangle_R$, but for convenience these can be chosen to match the $c\neq0$ cases. That is, 
\begin{align}
 |\Psi_{c=0}^{p}\rangle = \frac{1}{\sqrt{d}}\sum_{j=0}^{d/2-1} e^{i\phi_{j,p}} & \Bigl(|2j\rangle_L |2j\rangle_R \nonumber \\
 & +(-1)^p|2j+1\rangle_L | 2j+1\rangle_R\Bigr),
 \label{eqn:czerosymm}
\end{align}
with phases $\phi_{j,p}$ as in Equation \ref{eqn:phijp}. 

\noindent  \textit{The number of symmetric vs anti-symmetric states---}Given that the $c\neq0$ correlation classes is evenly split between exchange symmetric and antisymmetric states and the $c=0$ correlation class contains only symmetric states, the total number of symmetric states is $\frac{(d-1)d}{2}+d = \frac{(d+1)d}{2}$ and the number of antisymmetric states is $\frac{(d-1)d}{2}$. We note these numbers match those of the case $d = 2^n$ \cite{sych_complete_2009, pisenti_distinguishability_2011}---for example, for $d=2$ there are 3 symmetric and 1 antisymmetric states: the triplet ($|\Phi^{\pm}\rangle,|\Psi^+\rangle$) and singlet ($|\Psi^-\rangle$) Bell states.

\noindent  \textit{Example symmetric basis for $d=6$---}We illustrate our construction by explicitly finding exchange-symmetric Bell states for $d=6$. First we identify correlation classes. In general, there are many different ways to construct such a set satisfying our constraints. Choosing the pairings according to \cite{lucas_recreations_1883}, a sample round-robin tournament configuration of six players labeled $0$ to $5$ is:
\begin{align*}
\text{Round 1}&: [(0, 1), (2, 5), (3, 4)]\\
\text{Round 2}&: [(0, 2), (3, 1), (4, 5)]\\
\text{Round 3}&: [(0, 3), (4, 2), (5, 1)]\\
\text{Round 4}&: [(0, 4), (5, 3), (1, 2)]\\
\text{Round 5}&: [(0, 5), (1, 4), (2, 3)]\\
\end{align*}

To create the $c \neq 0$ correlation classes, we thus associate pairings $(s_{jc}, t_{jc})$ with basis states $|s_{jc}\rangle |t_{jc} \rangle$ and $|t_{jc}\rangle |s_{jc} \rangle$ to yield:

\begin{align}
 c &= 1: \; \{ |0\rangle |1\rangle , |1\rangle |0\rangle , |2\rangle |5\rangle , |5\rangle |2\rangle , |3\rangle |4\rangle , |4\rangle |3\rangle)\} \nonumber\\
 c &= 2: \; \{ |0\rangle |2\rangle , |2\rangle |0\rangle , |1\rangle |3\rangle , |3\rangle |1\rangle , |4\rangle |5\rangle , |5\rangle |4\rangle)\}\nonumber\\
 c &= 3: \; \{ |0\rangle |3\rangle , |3\rangle |0\rangle , |2\rangle |4\rangle , |4\rangle |2\rangle , |1\rangle |5\rangle , |5\rangle |1\rangle)\}\nonumber\\
 c &= 4: \; \{ |0\rangle |4\rangle , |4\rangle |0\rangle , |3\rangle |5\rangle , |5\rangle |3\rangle , |1\rangle |2\rangle , |2\rangle |1\rangle)\}\nonumber\\
 c &= 5: \; \{ |0\rangle |5\rangle , |5\rangle |0\rangle , |1\rangle |4\rangle , |4\rangle |1\rangle , |2\rangle |3\rangle , |3\rangle |2\rangle)\}.
\end{align}

Next we find the phase classes by constructing the discrete Fourier transform of size $\frac{6}{2} \times \frac{6}{2}$: 
\begin{equation}
 \phi_{j, p} = \left \lfloor{\frac{p}{2}} \right \rfloor \frac{2\pi j}{3}.
\end{equation}

Then Equations \ref{eqn:cnotzerosymm} and \ref{eqn:czerosymm} specify a complete orthonormal basis of $d=6$ Bell states. We display a few examples below, leaving off the L/R subscripts for visual simplicity:

\begin{align}
 |\Psi_0^4\rangle &=\frac{1}{\sqrt{6}}(|0\rangle |0\rangle + |1\rangle |1\rangle + e^{4 \pi i / 3}|2\rangle |2\rangle + e^{4 \pi i / 3}|3\rangle |3\rangle \nonumber\\
 & + e^{8 \pi i / 3}|4\rangle |4\rangle+ e^{8 \pi i / 3}|5\rangle |5\rangle), \nonumber\\
 |\Psi_0^5\rangle &= \frac{1}{\sqrt{6}}(|0\rangle |0\rangle - |1\rangle |1\rangle + e^{4 \pi i / 3}|2\rangle |2\rangle - e^{4 \pi i / 3}|3\rangle |3\rangle \nonumber\\
 & + e^{8 \pi i / 3}|4\rangle |4\rangle- e^{8 \pi i / 3}|5\rangle |5\rangle), \nonumber\\
 |\Psi_1^1\rangle &= \frac{1}{\sqrt{6}}(|0\rangle |1\rangle - |1\rangle |0\rangle + |2\rangle |5\rangle - |5\rangle |2\rangle+ |3\rangle |4\rangle - |4\rangle |3\rangle), \nonumber\\
 |\Psi_2^5\rangle &= \frac{1}{\sqrt{6}}(|0\rangle |2\rangle - |2\rangle |0\rangle + e^{4\pi i /3}|1\rangle |3\rangle - e^{4\pi i /3}|3\rangle |1\rangle \nonumber\\
 &+ e^{8\pi i /3}|4\rangle |5\rangle -e^{8\pi i /3} |5\rangle |4\rangle).
\label{eqn:newd6examples}
\end{align}
We can compare these states to those given by Equation \ref{eqn:generalbell} for the same correlation ($m = c$) and phase ($n = p$) classes:
\begin{align}
|\Psi_0^4\rangle &= \frac{1}{\sqrt{6}}(|0\rangle|0\rangle + e^{4 \pi i / 3}|1\rangle|1\rangle + e^{8 \pi i / 3}|2\rangle|2\rangle + e^{12 \pi i / 3}|3\rangle|3\rangle \nonumber \\
&+ e^{16 \pi i / 3}|4\rangle|4\rangle + e^{20 \pi i / 3}|5\rangle|5\rangle), \nonumber \\
|\Psi_0^5\rangle &=\frac{1}{\sqrt{6}}(|0\rangle|0\rangle + e^{5 \pi i / 3}|1\rangle|1\rangle + e^{10 \pi i / 3}|2\rangle|2\rangle + e^{5 \pi i }|3\rangle|3\rangle \nonumber \\
&+ e^{20 \pi i / 3}|4\rangle|4\rangle + e^{25 \pi i / 3}|5\rangle|5\rangle), \nonumber \\
|\Psi_1^1\rangle &= \frac{1}{\sqrt{6}}(|0\rangle|1\rangle + e^{i\pi/3}|1\rangle|2\rangle + e^{2\pi i/3}|2\rangle|3\rangle + e^{\pi i}|3\rangle|4\rangle \nonumber \\
&\quad + e^{4\pi i/3}|4\rangle|5\rangle + e^{5\pi i/3}|5\rangle|0\rangle), \nonumber \\
|\Psi_2^5\rangle &= \frac{1}{\sqrt{6}}(|0\rangle|2\rangle + e^{5\pi i/3}|1\rangle|3\rangle + e^{10\pi i/3}|2\rangle|4\rangle \nonumber \\
&\quad + e^{5\pi i}|3\rangle|5\rangle+ e^{20\pi i/3}|4\rangle|0\rangle + e^{25\pi i/3}|5\rangle|1\rangle).
\label{eqn:oldd6examples}
\end{align}

The entangled basis states of Equation \ref{eqn:newd6examples} are clearly eigenstates of left/right particle state exchange, whereas Equation \ref{eqn:oldd6examples} generated from the most common definition are not eigenstates of exchange.

\noindent  \textit{Connection to hyperentangled basis---}We can draw a connection between our basis construction and the existing exchanged-symmetrized construction for $d=2^n$ \cite{pisenti_distinguishability_2011, sych_complete_2009}. The general form of constructing the correlations remains the same, but since the authors there considered tensor products of $d=2$ Bell states, the phase factors between two-particle basis states were restricted to $\pm1$. This can be easily understood as a slight variation on our framework. For $d=2^n$, phase factors $e^{i \phi_{j, p}}$ can be restricted to the values $\pm 1$ by replacing the DFT with generalized Hadamard matrices, also known as Walsh matrices, which can be constructed via the recursive definition \cite{sylvester_lx_1867}:
\begin{equation}
 H(2^n) = H(2) \otimes H(2^{n-1}),
 \label{eqn:walsh}
\end{equation}
where $H(2)$ is the $2 \times 2$ Hadamard matrix, using a normalization convention in which entries have magnitude $1$.
Note that all the Walsh matrices therefore have mutually orthogonal columns composed of entries $\pm1$. Thus phase classes for $d=2^n$ Bell states can be constructed equally well using
\begin{equation}
e^{i\phi_{j, p}} = [H(2^{n-1})]_{j, p}.
\end{equation}

As an example, we consider $d=8$. We could by default use the DFT of size $4\times4$ to construct the phases:
\begin{equation}
\operatorname{DFT}(4) = \frac{1}{2}
\begin{bmatrix}
1 & 1 & 1 & 1 \\
1 & i & -1 & -i \\
1 & -1 & 1 & -1 \\
1 & -i & -1 & i
\end{bmatrix}.
\end{equation}
However, we could use equally well the Walsh matrix of size $4 \times 4$:
\begin{align}
 H(4) &= H(2) \otimes H(2)\\
 &= \begin{bmatrix}
 1 & 1 & 1 & 1\\
 1 & -1 & 1 & -1\\
 1 & 1 & -1 & -1\\
 1 & -1 & -1 & 1
 \end{bmatrix}.
\end{align}
$H(4)$, like $\operatorname{DFT}(4)$, has mutually orthogonal columns, but those of $H_4$ contain only $1$ and $-1$. Thus the Walsh matrices could be inserted seamlessly into the phase portion of the state construction procedure and would produce entangled qudit Bell states corresponding to the hyperentangled constructions \cite{pisenti_distinguishability_2011, sych_complete_2009}.

\textit{No symmetrized Bell basis for odd $d$---}The construction above relies on even single-particle Hilbert space dimension $d$; here we briefly establish that no Bell basis symmetrized with respect to particle exchange can exist for odd $d$. It is well known that every fully entangled bipartite pure state can be expressed in the form $|\Psi\rangle = \frac{1}{\sqrt{d}}\sum_{j=0}^{d-1} |\phi_j\rangle_L |\psi_j\rangle_R$ for some orthonormal bases $\{|\phi_j\rangle_L\}$ and $\{|\psi_j\rangle_R\}$. Thus when re-expressed in the form of Equation \ref{eqn:general_pbis}, every fully entangled bipartite state must have a unitary $d \times d$ coefficient matrix $\chi$. A complete basis of entangled states requires the existence of $d^2$ linearly independent coefficient matrices. Exchange (anti)symmetry corresponds to $\chi_{ij} = (-)\chi_{ji}$. However, every skew-symmetric matrix of odd dimension has determinant zero, and there are only $d(d+1)/2$ linearly independent symmetric matrices. Therefore no qudit Bell basis of exchange-symmetrized states can exist for odd $d$.

\noindent  \textit{Maximal distinguishability of bell states for even $d$---}We have established a general exchange-symmetrized Bell basis. We can exploit the symmetry properties of the basis to yield a simple proof of the maximum number of qudit Bell states that can be distinguished by a linear evolution and local measurement (LELM) device. Bell-basis measurements are required in many applications, including quantum teleportation \cite{bennett_teleporting_1993, sun_quantum_2016}, quantum repeaters \cite{briegel_quantum_1998, uphoff_integrated_2016}, and quantum dense coding \cite{bennett_communication_1992}. Despite their in-principle limitations, LELM devices remain relevant to numerous demonstrations of these quantum communication protocols because of their reliability in practice \cite{barreiro_beating_2008, van_houwelingen_quantum_2006, mirhosseini_high-dimensional_2015}. 
\begin{figure}
 \includegraphics[width=70mm]{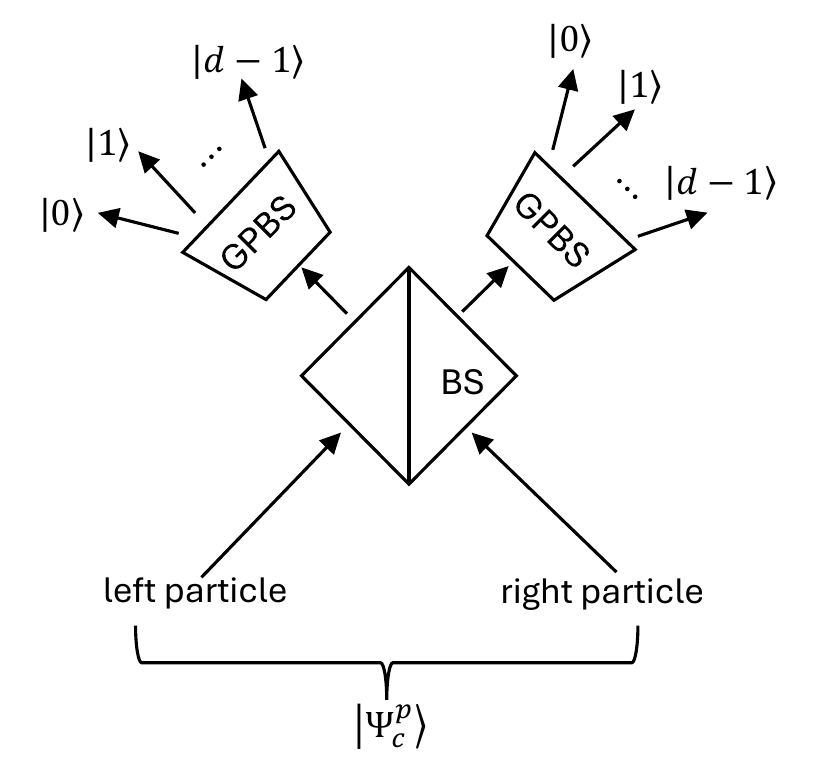}
 \caption{Optimal LELM device to distinguish qudit Bell states ($|\Psi^p_c\rangle$ as in Equations \ref{eqn:cnotzerosymm} and \ref{eqn:czerosymm}) of even single-particle dimension $d$. BS is a 50/50 beam splitter, and trapezoids represent generalized polarizing beam splitters (GPBS) that spatially separate single-particle basis states. The output channels shown couple to particle-number resolving detectors.}
 \label{fig:detector}
\end{figure}

Let the number of qudit Bell states unambiguously distinguishable under LELM to be $k$. Previously, only the results for $d=2, \, k=3$ \cite{lutkenhaus_bell_1999, vaidman_methods_1999}, $d=3, \, k=3$, and $d = 2^n$, $k=2^{n+1}-1$ \cite{pisenti_distinguishability_2011,wei_hyperentangled_2007} were known. Most relevant to our work is Ref. \cite{pisenti_distinguishability_2011}, which examined Bell states with single-particle dimension $d = 2^n$ and considered a hyperentangled basis containing only symmetric or antisymmetric states. That work demonstrated that, using an exchange-symmetrized Bell basis, the maximal distinguishability is
\begin{equation}
k = 2d - 1.
\label{eqn:maximalk}
\end{equation}

In the argument of Ref. \cite{pisenti_distinguishability_2011}, an upper bound of $2d-1$ LELM-distinguishable Bell-state classes is independent of the specific value of the single-particle Hilbert space dimension $d$.  The criteria of linear evolution and local measurement lead to detection signatures that consist of a pair of particle detections across $2d$ detection channels.  Maximal entanglement implies that any Bell state must be capable of triggering any detector at least once, meaning that for any ``first" particle detection there are only $2d$ available ``second" detections available to distinguish between Bell-state classes.  Finally, for indistinguishable particle pairs, the number of LELM-distinguishable Bell states is further decreased by one because of fundamental particle exchange symmetry:  for fermions no single detector may be triggered twice, and for bosons one of the detection signatures involving each detection channel $|i\rangle$ either does not occur or overlaps with the Bell states in another detection signature involving $|i\rangle$.  While Ref. \cite{pisenti_distinguishability_2011} presented this reasoning in detail for hyperentangled qubit Bell states with $d=2^n$, it applies equally for any $d$ provided the Bell basis is symmetrized under particle exchange.

 For qudit Bell states in our symmetrized basis, the optimal device can be realized experimentally with a 50-50 BS (beam-splitter) and then two pairs of GPBS (generalized polarizing beam splitters) of dimension $d$ creating $d$ output channels at each end of the GPBS, as shown in Figure \ref{fig:detector}. Such an apparatus generalizes the form of previous optimal distinguishability schemes for qubit pairs \cite{weinfurter_experimental_1994,braunstein_measurement_1995, michler_interferometric_1996, lutkenhaus_bell_1999, vaidman_methods_1999}. The main physical insight of Ref. \cite{pisenti_distinguishability_2011} is that this device exploits the symmetrized nature of the input states: by the generalized Hong-Ou-Mandel (HOM) effect, bipartite symmetric states will have both particles appear on the same output port of the beam splitter (bunching), whereas antisymmetric states will have one particle on each side (anti-bunching). Therefore, we can convert some amount of phase information---i.e., the parity of $p$---into spatial information.

This scheme does not consider the use of auxiliary modes.  However, it has been shown that the use of auxiliary modes of definite particle number cannot increase the number of distinguishable Bell states in an LELM device with projective measurement \cite{van_loock_simple_2004,carollo_role_2002}.

\textit{Application: Dense coding}---To give an example application of our basis to a specific quantum information procedure, we present an outline for a dense coding protocol for qudits using our Bell state distinguishability result for LELM devices. 
 
We suppose Alice and Bob use Bell states of an even $d$ to encode their states. They share the fully entangled state $|\Psi_0^0 \rangle$, and define a set of $2d-1$ codewords corresponding to one Bell state per distinguishability class. By analogy to the qubit case, the message that Alice transmits to Bob is encoded in which of these states $|\Psi_c^p \rangle$ she creates via a local unitary operation on her qudit, i.e. $U_A \otimes I_B$. After carrying out her chosen local operation, Alice sends her qudit to Bob, who can reliably distinguish between all $2d-1$ codewords using the device in Figure \ref{fig:detector}. 

For instance, Alice and Bob may choose their $2d-1$ codewords to be the $c=0$, $p=0$ and  $c\neq 0$, $p=0,1$ states. Alice can transform an initially shared $c=0$, $p=0$ state to any of the others by pairwise swapping her single-qudit basis states to achieve the desired $c$ and, to make $p=1$, flipping the phases of an appropriate 50\% of them. For example, for $d=6$ suppose Alice wishes to send the codewords $c=1$, $p=0$ or $c=1$, $p=1$. To transform $|\Psi_0^0\rangle \to |\Psi_1^0\rangle$, Alice performs
\begin{align}
    |0\rangle \to |1\rangle \; \; \; & \; \; \; |1\rangle \to |0\rangle \nonumber \\
    |2\rangle \to |5\rangle \; \; \; & \; \; \; |5\rangle \to |2\rangle \nonumber \\
    |3\rangle \to |4\rangle \; \; \; & \; \; \; |4\rangle \to |3\rangle.
\end{align}
 To create $|\Psi_1^1\rangle$, Alice performs
\begin{align}
    |0\rangle \to -|1\rangle \; \; \; & \; \; \; |1\rangle \to |0\rangle \nonumber \\
    |2\rangle \to -|5\rangle \; \; \; & \; \; \; |5\rangle \to |2\rangle \nonumber \\
    |3\rangle \to -|4\rangle \; \; \; & \; \; \; |4\rangle \to |3\rangle.
\end{align}

\noindent \textit{Conclusion---}In qudit pairs of even single-particle Hilbert space dimension $d$, we have introduced a Bell basis composed of states with definite symmetry under exchange of the internal states of the particles in the two distinct spatial channels (L,R). We have shown that an LELM device can unambiguously distinguish $2d-1$ of these exchange-symmetrized qudit Bell states. Furthermore, we have sketched a qudit dense coding protocol with channel size $2d-1$ with these states.

It remains an open question whether the same maximal distinguishability bound is achievable for unsymmetrized Bell states, as symmetrized and unsymmetrized bases are not, in general, equivalent under local unitary transformations. 

\noindent \textit{Acknowledgments---}The authors thank Ben Hartley, who contributed to early stages of this project. O.S. also thanks Ami Radunskaya, Thomas Moore, and Jorge Moreno for constructive comments on this work, Dorian Gangloff and Apollo Matsoukas for insights on presenting the material, Sathyawageeswar Subramanian for suggestions of connections between exchange symmetry and computing, and Robert Fricke, Tom Tang, Chris Wang, Larry Liu for helpful discussions. We thank an anonymous referee for their helpful comments.

\input{output_revised.bbl}


\end{document}

%% file: output_revised.bbl
%